# Thermodynamic properties for metal oxides from first-principles

Joakim Brorsson,[1] Ivana Staničić,[2] Jonatan Gastaldi,[2] Tobias Mattison,[2] and Anders Hellman[1, *]

[1]*Chalmers University of Technology, Department of Physics, SE-412 96 Gothenburg, Sweden*
[2]*Chalmers University of Technology, Department of Space, Earth and Environment, SE-412 96 Gothenburg, Sweden*

In this study, an efficient first-principles approach for calculating the thermodynamic properties of mixed metal oxides at high temperatures is demonstrated. More precisely, this procedure combines density functional theory and harmonic phonon calculations with tabulated thermochemical data to predict the heat capacity, formation energy, and entropy of important metal oxides. Alloy cluster expansions are, moreover, employed to represent phases that display chemical ordering as well as to calculate the configurational contribution to the specific heat capacity. The methodology can, therefore, be applied to compounds with vacancies and variable site occupancies. Results are, moreover, presented for a number of systems of high practical relevance: Fe–K–Ti–O, K–Mn–O, and Ca–Mn–O. In the case of ilmenite ($FeTiO_3$), the agreement with experimental measurements is exceptionally good. When the generated data is used in multi-phase thermodynamic calculations to represent materials for which experimental data is not available, the predicted phase-diagrams for the K–Mn–O and K–Ti–O systems change dramatically. The demonstrated methodology is highly useful for obtaining approximate values on key thermodynamic properties in cases where experimental data is hard to obtain, inaccurate or missing.

## I. INTRODUCTION

Thermodynamic calculations are widely applied to understand and predict thermochemical processes; they have proven to be especially useful when considering ashrelated problems during fuel conversion, including combustion [1], gasification [2–4], and chemical-looping [5–7]. In addition to providing faster results compared to experimental measurements, this approach can flag for issues that might be encountered in practical applications. A major limitation, however, is the availability of experimental data. In particular, inaccurate results will be obtained from this type of multi-component, multiphase equilibrium calculations if the databases used do not include all stable compounds. This is a significant obstacle in many present-day thermochemical conversion technologies, such as gasification and chemical-looping processes, for instance because the combination of active materials and ashes from biomass can lead to complex side-reactions [1, 6–8]. This is especially true for chemical looping combustion (CLC), in which oxygen carriers (OCs) are used to transport oxygen from the air to the fuel. A concrete example is the interaction between the benchmark OC ilmenite with potassium. This is regarded as one of the most problematic ash element in CLC of biomass (as well as other thermal conversion systems) and is, for example, known to cause problems in boilers due to the formation of silicates with low melting temperatures. Researchers have previously found that the ilmenite captures potassium when used as an OC [9, 10] and, specifically, attributed this phenomenon to the formation of the experimentally observed $KTi_8O_{16}$ phase. Unfortunately, the thermodynamic databases of relevant K–Ti–O compounds have, until now, been incomplete, making it impossible to simulate this system accurately. In cases such as this, where little or no thermodynamic data is available, a potential solution is to generate the required information using first-principles calculations. Indeed, such an approach could be viable in situations when measurements are difficult to perform or inherently inaccurate because of systematic errors related to the, e.g., phase behaviour of the material. By combining the calculated thermodynamic properties with data from existing databases, it is possible to investigate the stability of the missing compounds for the purpose of either predicting or explaining experimental observations. With regards to ash elements, for example, this allows the prediction of problematic properties and elucidation of potential remedies, as well as a better understanding of their interactions with different OCs, thereby providing guidance for the selection of the latter.

Recently, a semi-empirical methodology for estimating the thermodynamic properties of mixed oxides was proposed by Benisek and Dachs [11] and, moreover, shown to provide a reasonable agreement with experiments for a number of minerals under relevant conditions. In particular, it relies on a combination of reaction enthalpies, at 0K, calculated with density functional theory (DFT) and estimates of heat capacities, as functions of temperature, based on the harmonic approximation, as well as tabulated experimental data for monometallic oxides. The aim of this work is to utilize a similar procedure for predicting thermodynamic parameters for complex metal oxides, of relevance for chemical-looping with biomass. Specifically, a modified version of the aforementioned methodology is implemented that also takes chemical ordering into account, via the construction of alloy cluster expansions (CEs). Since this computational methodology is applicable to any metal oxide, and even other stable crystalline materials, it is believed to be useful for estimating high temperature thermodynamic properties of solids in a broad range of different contexts. Even so, it should be emphasized that the selected phases have all been found during chemical-looping processes. The latter represent breakthrough technologies for thermochemical conversion of fuels and their commercialization is regarded as key to meeting the climate targets because of the inherently low emissions as well as possibilities for carbon capture. Specifically, results from several systems of relevance for CLC are presented, i.e. K–Mn–O, Ca–Mn–O, and K–Fe–Ti–O.

## II. METHOD

In complex higher order systems, multicomponent thermodynamic calculations can generate stable phases based on minimizing the Gibbs free energy. Each component must, more precisely, be described by several state functions, here in the form of the standard enthalpies of formation and entropies, as well as the temperature variation of the heat capacity. In this work, these are obtained from first-principles calculations based on a modified version of a methodology originally proposed by Benisek and Dachs [11] Specifically, this approach is applied to a number of complex K–Mn–O, Ca–Mn–O, and K–Fe–Ti–O oxides, namely $KMn_2O_4$, $K_2MnO_4$, and $K_3MnO_4$ together with $CaMn_{14}SiO_{24}$ and $Ca_{19}Mn_2(PO_4)_{14}$, as well as $K_{0.4}Fe_{0.4}Ti_{0.6}O_2$, $K_{0.85}Fe_{0.85}Ti_{0.15}O_2$, $KTi_8O_{16}$, $KTi_8O_{16.5}$ (represented by $KTi_8O_{12}$), and $FeTiO_3$, where the later was included as a reference. Even though a comprehensive literature review has revealed that the selected compounds have been shown to form under experimental conditions, these observations cannot be verified or explained via simulations due to a lack of thermodynamic data [6, 7]. Hence, these systems are ideally suited for testing the procedure outlined below.

### A. Procedure and Theory

As was shown by Benisek and Dachs[11], the following procedure (see Figure 1), which is described for $FeTiO_3$, can be used to calculate the thermodynamic properties of any metal oxide.

1. Identify a possible chemical reaction that describes the formation of the compound of interest from a set of binary oxides, i.e.,

$$FeO + TiO_2 \leftrightarrow FeTiO_3. \quad (1)$$

2. Compute the heat capacity for the reaction at constant volume,

$$\Delta_R C_V = C_{V,FeTiO_3} - (C_{V,FeO} + C_{V,TiO_2}). \quad (2)$$

3. Assume that the heat of reaction at 0K can be represented by the corresponding change in the internal energy,

$$\Delta_R H_{0K} = \Delta_R U_{0K} + P\Delta_R V_{0K} \approx \Delta_R U_{0K}$$
$$= U_{FeTiO0K3} - \left(U_{FeO}^{0K} + U_{TiO_2}^{0K}\right), \quad (3)$$

which is based on the assumption that the volume term $P\Delta_R V_{0K}$, which relates the enthalpy and internal energy, can be neglected.

4. Estimate the enthalpy and entropy of the reaction at room temperature via:

$$\Delta_R H^{298.15\,K} \approx \Delta_R H^{0\,K} + \int_0^{298.15} \Delta_R C_V\, dT, \quad (4)$$

$$\Delta_R S^{298.15\,K} \approx \Delta_R S^{0\,K} + \int_0^{298.15} \frac{\Delta_R C_V}{T}\, dT. \quad (5)$$

As Benisek and Dachs demonstrate, the error that results from using the reaction heat capacity at constant volume rather than pressure is relatively small. It should also be noted that a presumed benefit of the above formula, compared to calculating the entropy and formation enthalpy at 0K before adding the contribution from the heat capacity up to 293.15K, is that the magnetic contributions for the product and the reactants will, at least partially, cancel. Nevertheless, this means that it is necessary to compute not only $C_{V,FeTiO_3}$ but also $C_{V,FeO}$ and $C_{V,TiO_2}$.

5. Use data for the monometallic oxides, in this case, FeO and $TiO_2$ from the NIST-JANAF thermochemical tables [12], to correct the formation enthalpy and entropy of the multinary compound:

$$\Delta_f H^{298.15\,K} = \Delta_R H^{298.15\,K} + 2\Delta_f H^{298.}_{FeO15K}$$
$$+ \Delta_f H_{TiO298.215K}, \quad (6)$$

$$S_{298.15K} = \Delta_R S_{298.15K} + S_{FeO298.15K}$$
$$+ S_{TiO298.215K}. \quad (7)$$

A drawback with the approach outlined above is that one does not directly obtain the heat capacity above room temperature. Still, it is possible to derive an equation from which it can be estimated. The first step is to set the following alternative expressions for the enthalpy of formation equal:

$$\begin{cases} \Delta_f H^T &= \Delta_f H^{0K} + \int_0^T C_P\, dT' \\ \Delta_f H^T &\approx \Delta_R H^{0K} + \int_0^T \Delta_R C_V\, dT' \\ &\quad + \Delta_f H^T_{FeO} + \Delta_f H^T_{TiO_2} \end{cases}, \quad (8)$$

which gives the following result,

$$\int_0^T C_P\, dT' \approx \Delta_R H^{0\,K} + \int_0^T \Delta_R C_V\, dT'$$
$$+ \Delta_f H^T_{FeO} + \Delta_f H^T_{TiO} - \Delta_f H^{0\,K}. \quad (9)$$

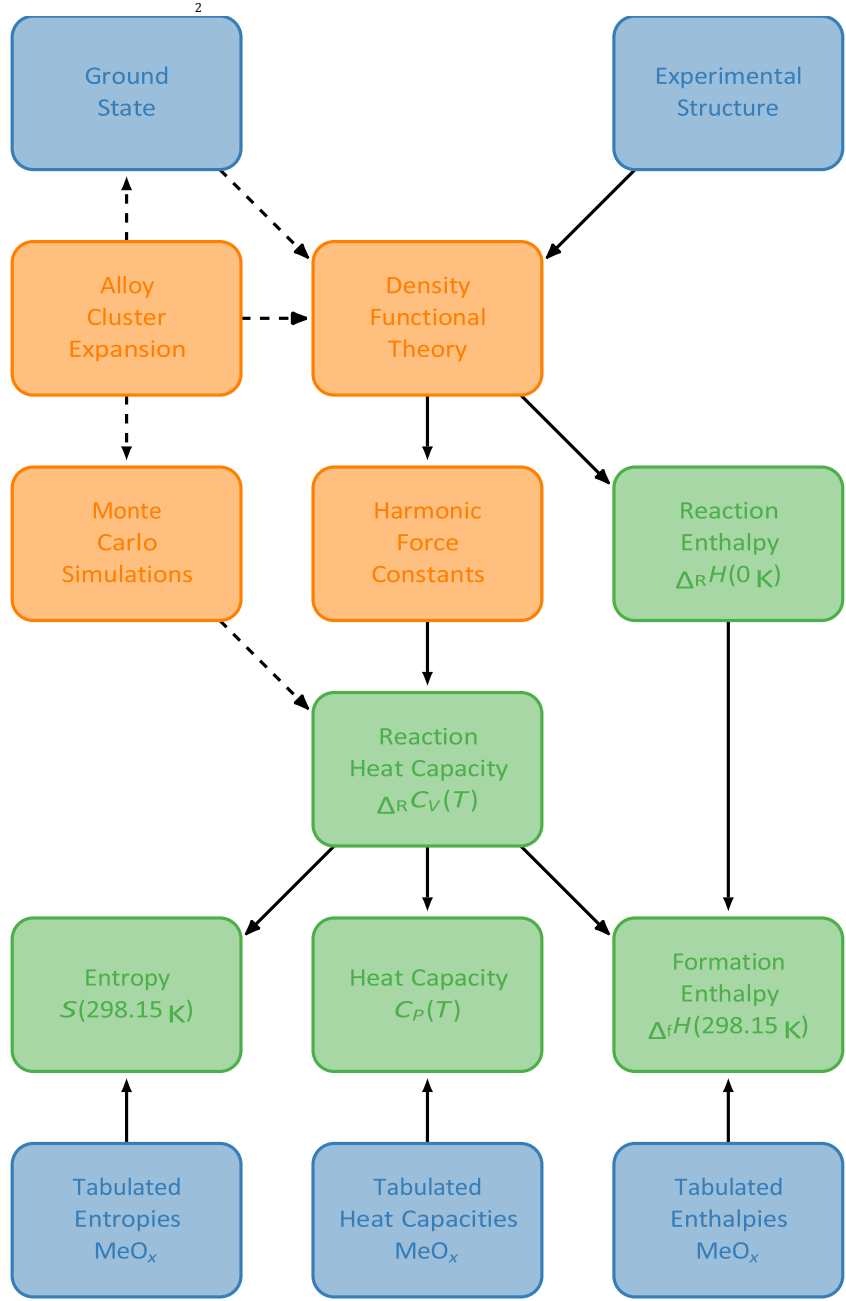

FIG. 1. Schematic illustration of the procedure for estimating the entropy, formation enthalpy, and heat capacity using a combination of first-principles calculations and tabulated experimental data for monometallic oxides.

Since the heat capacity at constant pressure is available for the binary oxides from the NIST-JANAF thermochemical tables, it is possible to express the corresponding formation enthalpies as,

$$\Delta_{\mathrm{f}} H_{\mathrm{X}}^{T} = \Delta_{\mathrm{f}} H_{\mathrm{X}}^{0} + \int_{0}^{T} C_{P,\,\mathrm{X}}\, dT', \tag{10}$$

where X = FeO, TiO$_2$. By inserting this expression into Equation 9 one obtains a relatively simple formula for the heat capacity of the compound of interest,

$$\int_{0}^{T} C_{P}\, dT' \approx \Delta_{\mathrm{R}} H^{0\,\mathrm{K}} - \Delta_{\mathrm{f}} H^{0\,\mathrm{K}} + \int_{0}^{T} \Delta_{\mathrm{R}} C_{V}\, dT'$$
$$+ 2\Delta_{\mathrm{f}} H_{\mathrm{FeO}}^{0} + \int_{0}^{T} C_{P,\,\mathrm{FeO}}\, dT'$$
$$+ \Delta_{\mathrm{f}} H_{\mathrm{TiO}_2}^{0} + \int_{0}^{T} C_{P,\,\mathrm{TiO}_2}\, dT'$$
$$\Rightarrow C_{P}(T) = \frac{d}{dT} \int_{0}^{T} C_{P}\, dT'$$
$$\approx \Delta_{\mathrm{R}} C_{V}(T) + C_{P,\,\mathrm{FeO}}(T)$$
$$+ C_{P,\,\mathrm{TiO}_2}(T). \tag{11}$$

In this way, the difference between the heat capacities at constant volume and pressure will, to some extent, be taken into account, similarly to the magnetic contribution.

It is important to note that FactSage requires that the temperature dependence of the specific heat at constant pressure is provided as a sum of power functions with at most eight terms,

$$C_P(T) = \sum_{i=1}^{8} k_i T^{p_i}. \tag{12}$$

It is, therefore, problematic to model certain features, such as the peaks typically observed at phase transitions [13], and, consequently, less relevant to include contributions from, for instance, chemical and magnetic ordering. Though various empirical formulas for the specific heat of minerals have been proposed [14], we will, for the sake of convenience, fit expressions with the same functional form as the one used in FactSage for most species, including $FeTiO_3$,

$$C_P(T) = k_0 + k_1 T^{-2} + k_2 T^{-0.5} + k_3 T^{-3}. \tag{13}$$

### B. Selection of Reference Data

The first step of the procedure detailed in the previous section is to identify a chemical formula that describes the formation of the target compound from a set of binary oxides, which have, when possible, been selected from the NIST-JANAF thermochemical tables [12]. It was, however, necessary to use data from other sources for a few compounds, namely $MnO$ [15] and $Mn_2O_3$ [16]. Since the properties of interest are state functions, all such combinations, if theoretically plausible, will be equally viable from a thermodynamic perspective. Even so, it is convenient to use the smallest possible subset of reactants that do not undergo any phase transformations, and for which data is available, within the relevant temperature interval. Based on these criteria, the following reactions were deemed to be the most suitable for the studied systems:

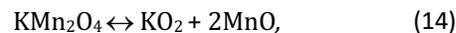

$$KMn_2O_4 \leftrightarrow KO_2 + 2MnO, \tag{14}$$

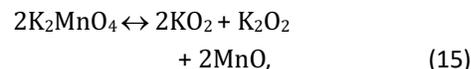

$$2K_2MnO_4 \leftrightarrow 2KO_2 + K_2O_2 + 2MnO, \tag{15}$$

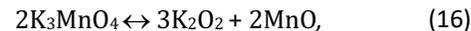

$$2K_3MnO_4 \leftrightarrow 3K_2O_2 + 2MnO, \tag{16}$$

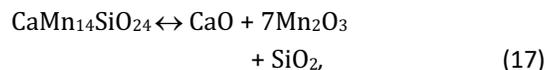

$$CaMn_{14}SiO_{24} \leftrightarrow CaO + 7Mn_2O_3 + SiO_2, \tag{17}$$

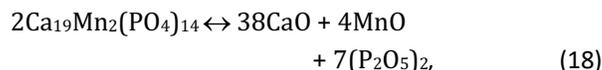

$$2Ca_{19}Mn_2(PO_4)_{14} \leftrightarrow 38CaO + 4MnO + 7(P_2O_5)_2, \tag{18}$$

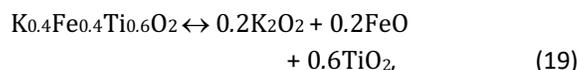

$$K_{0.4}Fe_{0.4}Ti_{0.6}O_2 \leftrightarrow 0.2K_2O_2 + 0.2FeO + 0.6TiO_2, \tag{19}$$

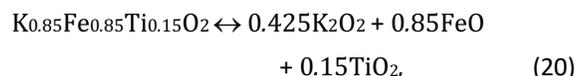

$$K_{0.85}Fe_{0.85}Ti_{0.15}O_2 \leftrightarrow 0.425K_2O_2 + 0.85FeO + 0.15TiO_2, \tag{20}$$

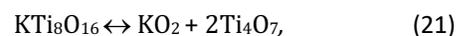

$$KTi_8O_{16} \leftrightarrow KO_2 + 2Ti_4O_7, \tag{21}$$

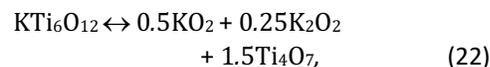

$$KTi_6O_{12} \leftrightarrow 0.5KO_2 + 0.25K_2O_2 + 1.5Ti_4O_7, \tag{22}$$

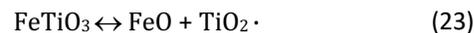

$$FeTiO_3 \leftrightarrow FeO + TiO_2. \tag{23}$$

### C. Representative Structures via Enumeration

For most of the products and reactants listed in the previous subsection, the lowest energy structures available from the Materials Project [17] database were used as starting points for the first-principles calculations. No entries could be found for $CaMn_{14}SiO_{24}$, $KTiO_8O_{16}$, $KTi_6O_{12}$, $K_{0.4}Fe_{0.4}Ti_{0.6}O_2$, and $K_{0.85}Fe_{0.85}Ti_{0.15}O_2$. A partial explanation is that all structures, except the first, contain vacancies and, in the last two cases, mixed occupancies on the Fe/Ti sites. Consequently, a different approach had to be employed. As a first step, experimentally determined crystal structures were retrieved from Springer Materials [18–22]. Although an entry for $Ca_{19}Mn_2(PO_4)_{14}$ does exist, the calculations were based on crystallographic information found in the Inorganic Crystal Structure Database (ICSD) [23, 24].

Fortunately, $KTi_8O_{16}$ represents a particularly simple case since the primitive unit cell contains a single vacancy, resulting in just two symmetry equivalent alternatives. In the case of $KTi_6O_{12}$, all possibilities were identified via enumeration using functionalities from the icet [25] toolkit. Because of the large number of configurations thus obtained, a filtering procedure had to be implemented. In particular, the ground state was taken as the lowest energy structure, after relaxation, among those with at least tetragonal symmetry (international tables of crystallography number above 75), and physically reasonable interatomic distances (>2.0Å). This procedure was repeated for $K_{0.4}Fe_{0.4}Ti_{0.6}O_2$ as well, although based on slightly different conditions. In addition to requiring that the symmetry was orthorhombic or higher (international tables of crystallography number above 15) it was necessary to limit the aspect ratio (< 10) as well. The latter condition was, more precisely, necessary because the structures with extremely elongated supercells were not only deemed unphysical but also proved difficult to relax or were found to correspond to relatively high energies.

### D. Ground State from Cluster Expansions

While the enumeration method described above can be applied to $K_xFe_xTi_{1-x}O_2$ with $x = 0.4$, it fails when considering $x = 0.15$ due to the vast configurational space resulting from the mixed occupancies on the Fe/Ti and K/vacancy sites. To circumvent this problem, alloy Ces, as implemented in the icet toolkits [26], were constructed for both compositions since such models allow the use of efficient methods for identifying ground state configurations. This means that the property of interest, here the energy, is expanded as a sum over subsets of atoms, often referred to as clusters, to an arbitrary order. In practice, however, it is necessary to consider the dimensions of the resulting system of equations, or fit matrix. Specifically, it is generally advisable to limit the number of terms so that these do not exceed the size of the training data. Otherwise, there will be fewer rows than columns, which means that the system is underdetermined and therefore more difficult to solve. In addition, it is necessary to ensure that the condition number, which gives a measure of the sensitivity of the solution with respect to changes in the input data, remains reasonable. Fortunately, the nature of interatomic interactions means that the solution is generally sparse and that the relative influence of the individual terms decreases with both order and distance. For this reason and because the models would only need to represent a small portion of the entire phase space for the cases at hand, i.e., a single composition, it was deemed possible to only consider terms up to the third order. Since this is not enough to truncate the sum, cutoffs for pairs and triples had to be selected as well. Depending on the fit method, this choice is also influenced by the root mean square error (RMSE), which should be kept as low as possible. As a first step of the CEconstruction, 120 $K_{0.85}Fe_{0.85}Ti_{0.15}O_2$ training structures were generated through random occupation of the sites with mixed occupancies. The same procedure was repeated for $K_{0.4}Fe_{0.4}Ti_{0.6}O_2$, based on the 20 high symmetry configurations identified using the procedure described in the previous subsection together with an arbitrary set of 98 enumerated structures. This was followed by extensive testing, which revealed that longer pairs could be allowed for $K_{0.4}Fe_{0.4}Ti_{0.6}O_2$ (7.5Å) compared to $K_{0.85}Fe_{0.85}Ti_{0.15}O_2$ (5.5Å) while the same limit was used for the triplets (4.5Å). After having relaxed the training structures, as described below, the Ces were fitted using two alternative methods, namely automatic relevance detection regression (ARDR) and ordinary least squares (OLS) with recursive feature elimination (RFE). By comparing the cross-validation RMSE scores, the former was deemed to yield the best CE, which was therefore used for training the final model. Finally, the ground states were identified through a procedure based on multiple integer programming [27], which is implemented in icet [25].

### E. Monte Carlo Sampling of Cluster Expansion

A benefit of having constructed alloy CEs is that they can be used to estimate the contribution to the heat capacity from chemical ordering. This was achieved via Monte Carlo (MC) simulations, as implemented in the mchammer module of icet [25]. Specifically, these were performed by starting from randomly occupied $3\times2\times1$ $K_{0.4}Fe_{0.4}Ti_{0.6}O_2$ and $2\times1\times1$ $K_{0.85}Fe_{0.85}Ti_{0.15}O_2$ supercells, respectively, and successively lowering the temperature from 2000K to 0K at a rate of 100K per 32000 MC cycles. From the recorded values on the variation of the energy, $E$, the configurational heat capacity was calculated [28] as $\Delta C_V^{conf} = \langle \Delta E \rangle / k_B T$, where $k_B$ is the Boltzmann constant and $T$ the absolute temperature while $\langle \cdot \rangle$ represents an ensemble average.

### F. Electronic Structure Calculations

For all representative structures, both the ionic positions and the cell metric were relaxed via DFTcalculations using the projector augmented wave (PAW) method [29] as implemented in the Vienna ab initio simulation package (VASP) [30]. The input files for the calculations were generated with the help of the pymatgenmaterials science library [31], to ensure that the settings were the same as in the Materials Project [17]. This meant employing a combination of generalized gradient approximations (GGA) [32] functionals, as parameterized by Perdew, Burke, and Ernzerhof (PBE) [33], together with an energy cutoff of 520eV. For the initial relaxation, the energy difference was required to be lower than $5 \times 10^{-5}$ eV and $5 \times 10^{-4}$ eV per atom for electronic and ionic convergence, respectively. To obtain more accurate energies and forces, this was followed by a re-relaxation step after fixing the former and latter parameter to $1 \times 10^{-6}$ eV and $1 \times 10^{-5}$ eV respectively, regardless of the number of atoms per unit cell. Because of convergence issues, these values had to be set an order of magnitude higher for $FeTiO_3$ and $K_{0.85}Fe_{0.85}Ti_{0.15}O_2$. Note that the same settings were used when calculating the forces for the rattled structures, generated as described in the next subsection. To account for strong correlation effects, certain systems were treated within the density functional theory with Hubbard correction (DFT+U) framework. In addition, corrections, which have been shown to provide significantly higher accuracy, were applied in order to minimize the impact of the inherent errors resulting from mixing of GGA and GGA+U [34, 35] as well as the poor representation of certain anionic species. Spin polarisation was also considered; the initial guess for the magnetic configuration was based on the information in the Materials Project database, for the structures retrieved therefrom, and otherwise assumed to correspond to a high spin ferromagnetic state. In addition, the tetrahedron method with Blöchl correction was employed to sample the reciprocal grid consisting of 1100 $k$-points per atom, which is slightly higher than the default (1000). The latter was, more precisely, generated based on the Monkhorst-Pack method except for systems with hexagonal symmetry, for which Γ-centered meshes were used instead. One should expect that the numerical convergence with respect to the total energy should be lower than 15meV per atom since this was reportedly achieved when using similar settings yet for a grid density of just 500 $k$-points per atom[17].

### G. Phononic Heat Capacity

The phonopy [36] package was used to determine the harmonic phonon contribution to the heat capacity, based on force constants (FCs) constructed with the help of the hiphive [26] code. As a first step, five "rattled" structures were generated per system, obtained by randomly displacing all atoms in a given supercell by a distance drawn from a Gaussian distribution with a 0.02Å standard deviation. After calculating the corresponding forces with VASP, as described in the previous subsection, the resulting dataset was used to fit a third-order force constant potentials (FCPs) that included terms up to the third order. Finally, the second-order FCs extracted from the latter was employed to compute the heat capacity, entropy, and free energy between 10K and 2500K, based on a uniform mesh with $5 \times 10^6$ $q$-points per reciprocal atom. The reason for including third-order terms is that this is known to have a stabilizing effect on the fitting and was, therefore, expected to give a higher accuracy [37].

### H. Generation of Phase Diagrams

The results obtained from the first-principles calculations were integrated with commercial equilibrium databases, more specifically FactPS and FToxid, that form part of the FactSage toolkit. Thereafter, phase diagrams for the K–Mn–O, Ca–Mn–O, and K–Fe–Ti–O systems were generated using the 8.2 version of the latter software.

### III. RESULTS AND DISCUSSION

### A. Calculated Thermal Properties

The procedure outlined in the previous section was employed to estimate the enthalpy of formation and entropy at room temperature together with values on the coefficients in the expression for the heat capacity for $CaMn_{14}SiO_{24}$, $Ca_{19}Mn_2(PO_4)_{14}$, $KMn_2O_4$, $K_2MnO_4$, and $K_3MnO_4$ as well as $KTi_8O_{16}$, $KTi_8O_{16.5}$, $K_{0.4}Fe_{0.4}Ti_{0.6}O_2$, $K_{0.85}Fe_{0.85}Ti_{0.15}O_2$, and $FeTiO_3$ (Table I). Because of limitations in the experimental data for the binary oxides, the expression in Equation 13 was fitted from 293.15K to 2000K for $K_3MnO_4$, 1600K for $K_{0.4}Fe_{0.4}Ti_{0.6}O_2$; $K_{0.85}Fe_{0.85}Ti_{0.15}O_2$; and $FeTiO_3$, 1500K for $Ca_{19}Mn_2(PO_4)_{14}$; $K_2MnO_4$; $KMn_2O_4$; $KTi_8O_{16}$; and $KTi_8O_{16.5}$, and 1400K for $CaMn_{14}SiO_{24}$.

An exceptionally good agreement is found between the results obtained for $FeTiO_3$ and the corresponding values from FactSage (Figure 2). The same holds true if compared with the data reported by Anovitz et al. [38], which is based on an alternative expression for the heat capacity,

$$C_P(T) = k_0' + k_1'T + k_2'T^2 + k_3'T^{-0.5} + k_4'T^{-2}. \quad (24)$$

This should be seen as clear evidence for the reliability of our estimates. It is worth noting the average error for the formation energies retrieved from the Open Quantum Materials Database (OQMD), which have been calculated using a procedure comparable to the one used in this study, is reportedly lower than the variations in the corresponding experimental data [39]. Indeed, the maximum error when comparing the calculated heat capacity, enthalpy of formation, and entropy with the data from FactSage (Anovitz et al.) are as small as 2.8Jmol$^{-1}$ K$^{-1}$ (6.6Jmol$^{-1}$ K$^{-1}$), 0.89kJmol$^{-1}$ (2.3kJmol$^{-1}$) and 1.6Jmol$^{-1}$ K$^{-1}$ (2.0Jmol$^{-1}$ K$^{-1}$), respectively. Such a high level of compliance is quite remarkable in light of the fact that the error in high temperature heat capacity measurements can be an order of magnitude larger (±10%) [40].

As should be expected, the properties of the various systems differ substantially in magnitude (see Table I and Figure 3). More importantly, there are significant variations between structurally similar compounds, indicating not the relevance of this study but also that the applied methodology is precise enough to account for minor differences in structure or composition. While it is not surprising that $Ca_{19}Mn_2(PO_4)_{14}$, which is structurally more complex than $CaMn_{14}SiO_{24}$, has a twice as high heat capacity and formation enthalpy, it is intriguing that the same holds true for $KTi_8O_{16}$ compared to $KTi_8O_{16.5}$. The difference between $K_{0.4}Fe_{0.4}Ti_{0.6}O_2$ and $K_{0.85}Fe_{0.85}Ti_{0.15}O_2$ is smaller but still significant. The situation is more complicated for the K–Mn–O system since the enthalpy is relatively low for $K_2MnO_4$ compared to $KMn_2O_4$ and $K_3MnO_4$, while the comparison of the heat capacities reveals that $K_3MnO_4 > KMn_2O_4 > K_2MnO_4$.

As was mentioned in the previous section, alloy CEs were constructed in order to find the ground states of $K_{0.4}Fe_{0.4}Ti_{0.6}O_2$ and $K_{0.85}Fe_{0.85}Ti_{0.15}O_2$. This also allowed us to calculate the entropic contribution from the chemical disorder on the K/vacancy and Fe/Ti sites. As expected, the phononic heat capacity is found to be an order of magnitude higher than the configurational part for both K$_{0.4}$Fe$_{0.4}$Ti$_{0.6}$O$_2$ and K$_{0.85}$Fe$_{0.85}$Ti$_{0.15}$O$_2$ (see Figure 4). Interestingly, the results also indicate the existence of an order-disorder transition, at around 400K, for the latter composition, but not the former. Thanks to a more limited phase space, it was possible to approximate the K$_{0.4}$Fe$_{0.4}$Ti$_{0.6}$O$_2$ ground state by the most stable configuration with the highest symmetry. Since the predicted thermal properties for this structure only differ minutely from those of the true ground state, it can be concluded that the construction of CEs is not strictly necessary in order to obtain good enough estimates. If enumeration of all configurations is not possible, however, it is difficult, if not impossible, to identify a suitable representative structure without a CE. Consequently, this construction could prove invaluable if considering complex materials with chemical disorders on multiple sublattices.

TABLE I. Thermodynamic properties in the form of the formation enthalpies ($\Delta_f H^{298.15K}$) and entropies ($S^{298.15K}$) at 298.15K, calculated via Equation 6 and Equation 7 respectively, together with the coefficients for the expression of the heat capacity in Equation 13.

| System | $\Delta_f H^{298.15K}$ (kJmol$^{-1}$) | $S^{298.15K}$ (Jmol$^{-1}$ K$^{-1}$) | $k_0$ (Jmol$^{-1}$ K$^{-1}$) | $k_1$ (JKmol$^{-1}$) | $k_2$ (Jmol$^{-1}$ K$^{-0.5}$) | $k_3$ (JK$^2$ mol$^{-1}$) |
|---|---|---|---|---|---|---|
| CaMn$_{14}$SiO$_{24}$ | $-8.3865 \times 10^6$ | 859.51 | 1776.9 | $4.3635 \times 10^7$ | $-23102$ | $-2.1352 \times 10^9$ |
| Ca$_{19}$Mn$_2$(PO$_4$)$_{14}$ | $-2.217 \times 10^7$ | 1464.7 | 4099.9 | $2.3553 \times 10^7$ | $-52548$ | $2.5942 \times 10^9$ |
| K$_3$MnO$_4$ | $-1.3847 \times 10^6$ | 185.26 | 648.08 | $7.5366 \times 10^7$ | $-14614$ | $-1.3289 \times 10^{10}$ |
| K$_2$MnO$_4$ | $-1.1936 \times 10^6$ | 183.84 | 324.12 | $1.2967 \times 10^7$ | $-4411.6$ | $-2.1655 \times 10^9$ |
| KMn$_2$O$_4$ | $-1.3551 \times 10^6$ | 179.29 | 256.07 | $6.9326 \times 10^5$ | $-1725.5$ | $-3.9211 \times 10^8$ |
| KTi$_8$O$_{16}$ | $-7.753 \times 10^6$ | 469.6 | 782.67 | $-1.6408 \times 10^7$ | $-3318.5$ | $1.4909 \times 10^9$ |
| KTi$_8$O$_{16.5}$ | $-5.9127 \times 10^6$ | 339.93 | 637.03 | $-3.2216 \times 10^6$ | $-4260.4$ | $-3.113 \times 10^8$ |
| K$_{0.4}$Fe$_{0.4}$Ti$_{0.6}$O$_2$ | $-8.5429 \times 10^5$ | 75.472 | 152.17 | $4.2947 \times 10^6$ | $-1771.6$ | $-7.0208 \times 10^8$ |
| K$_{0.4}$Fe$_{0.4}$Ti$_{0.6}$O$_2$ (GS) | $-8.5453 \times 10^5$ | 73.086 | 151.36 | $4.31 \times 10^6$ | $-1771.7$ | $-7.0488 \times 10^8$ |
| K$_{0.85}$Fe$_{0.85}$Ti$_{0.15}$O$_2$ (GS) | $-7.1722 \times 10^5$ | 86.347 | 221.19 | $1.9547 \times 10^7$ | $-3989.5$ | $-3.6146 \times 10^9$ |
| FeTiO$_3$ | $-1.2338 \times 10^6$ | 109.65 | 169.09 | $1.3492 \times 10^6$ | $-1175.8$ | $-3.4998 \times 10^8$ |
| FeTiO$_3$ (FToxid) | $-1.2331 \times 10^6$ | 108.63 | 150 | $-3.3237 \times 10^6$ | $-441.62$ | $3.4815 \times 10^8$ |

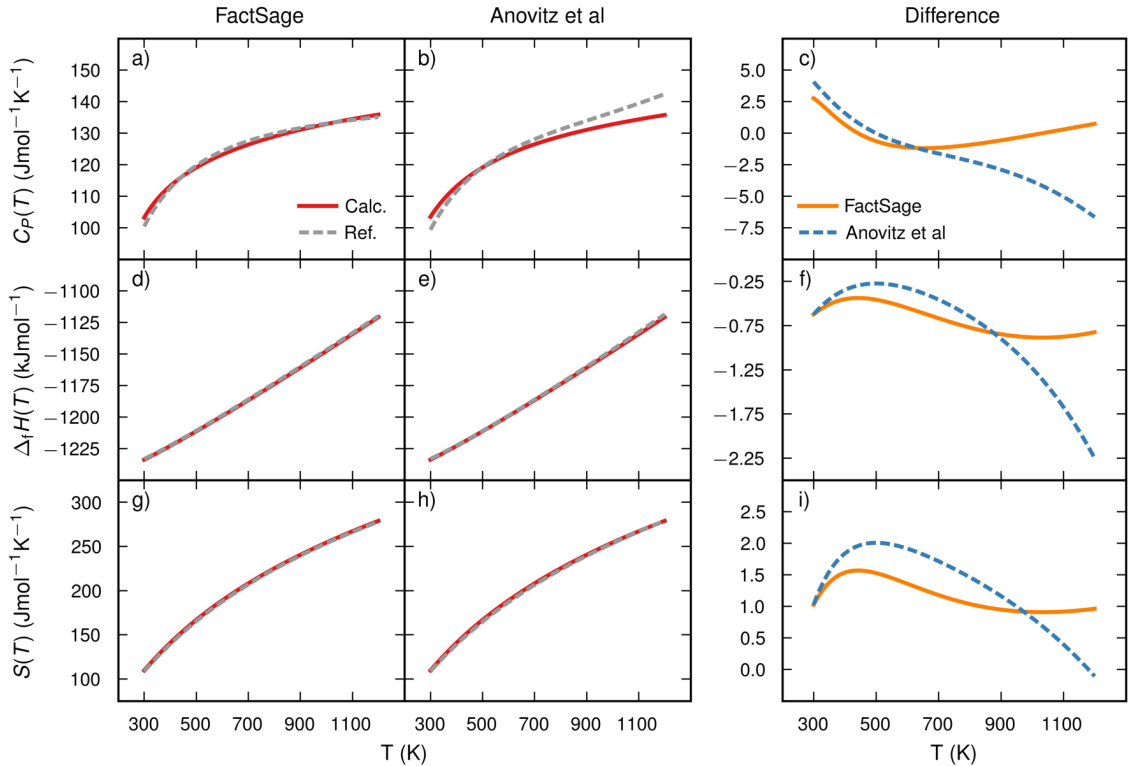

FIG. 2. Heat capacity (a-c), enthalpy of formation (d-f), and entropy (g-i) calculated (solid red line) based on the expressions used in FactSage (a, d, g) and Anovitz et al. [38] (b, e, h) together with the corresponding reference values (dashed grey line). The differences between the estimated and experimental data from FactSage (solid orange line) and Anovitz et al. (dashed blue line) are also shown (c, f, i).

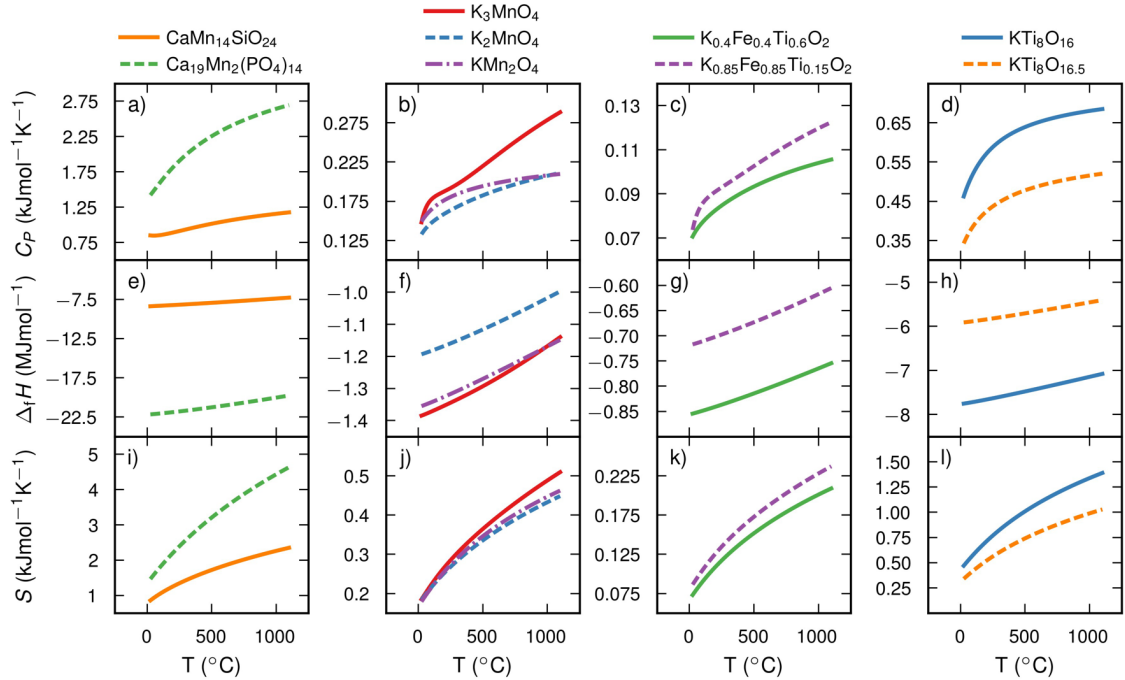

FIG. 3. Thermodynamic properties for CaMn$_{14}$SiO$_{24}$ (solid orange line in a, e, and i), Ca$_{19}$Mn$_2$(PO$_4$)$_{14}$ (dashed green line in a, e, and i), K$_3$MnO$_4$ (solid red line in b, f, and j), K$_2$MnO$_4$ (dashed blue line in b, f, and j), KMn$_2$O$_4$ (dash-dotted purple line in b, f, and j), K$_{0.4}$Fe$_{0.4}$Ti$_{0.6}$O$_2$ (solid green line in c, g, and k), K$_{0.85}$Fe$_{0.85}$Ti$_{0.15}$O$_2$ (dashed purple line in c, g, and k), KTi$_8$O$_{16}$ (solid blue line in d, h, and l), and KTi$_8$O$_{16.5}$ (dashed orange line in d, h, and l). This includes the heat capacity (a-d), fitted to an expression of the form used for FeTiO$_3$ in FactSage, as well as the enthalpy (e-h) and entropy (i-l).

### B. Impact on Phase Diagrams

Based on the phase diagrams generated using FactSage, it is clear that the data from the first-principles calculations have the largest impact on the K–Mn–O system (see Figure 5). In fact, KMn$_2$O$_4$, K$_2$MnO$_4$, and K$_3$MnO$_4$ are all stable at 950 °C and 1atm given that the partial oxygen pressure is sufficiently high ($p_{O_2} \gtrsim 10^{-11}$ atm). As should be expected, KMn$_2$O$_4$ dominates at low K/Mn ratio (K/(K+Mn) $\lesssim 0.35$). However, for a close to equal content of K and Mn, K$_3$MnO$_4$ also begins to form. With an increasing amount of Mn and a $p_{O_2}$ that approaches atmospheric levels, K$_2$MnO$_4$ starts to appear first together with KMn$_2$O$_4$ (K/(K+Mn) $\lesssim 0.7$) and then K$_3$MnO$_4$ (K/(K + Mn) $\lesssim 0.75$). At even lower K/Mn ratios, K$_3$MnO$_4$ is the most stable of the three phases, except at $p_{O_2}$ close to 1atm where it is replaced by a combination of K$_2$MnO$_4$ and either KO$_2$ or K$_2$O$_2$. Since significant fractions of K can be found in biomass, and residues therefrom, it is evident that the thermodynamic data generated in this study can have implications for thermal conversion systems using Mn-based OCs with bio-based fuels.

Although most of the Fe–K–Ti–O compounds have little or no influence on the corresponding phase diagrams at relevant conditions, the opposite is true for KTi$_8$O$_{16}$, which is predicted to be surprisingly stable. This is the most apparent when considering the K$_2$O–TiO$_2$ system (see Figure 6) at a low ratio between K and Ti (K$_2$O/TiO$_2$=0.1). In the absence of KTi$_8$O$_{16}$, K$_2$Ti$_6$O$_{13}$ is the dominating phase and, moreover, appears in combination with first rutile and then Ti$_x$O$_y$, with successively increasing Ti/O ratio, as the temperature becomes higher and oxygen pressure lower. In fact, it only disappears in favor of slag at very high temperatures and reducing conditions (log$_{10}$($p_{O_2}$) $\lesssim 0.03T - 53$). The behavior changes markedly when KTi$_8$O$_{16}$ is introduced. With increasing temperature and decreasing oxygen pressure, it first forms a mixture with K$_2$Ti$_6$O$_{13}$, replacing rutile, (log$_{10}$($p_{O_2}$) $\lesssim 0.026T - 38$) followed by slag in the region (log$_{10}$($p_{O_2}$) $\lesssim 0.053T - 76$) where various Ti$_x$O$_y$ phases would otherwise be observed. As has been previously reported [8], the inclusion of thermodynamic data for the aforementioned Fe–K–Ti–O phases directly influences simulations of the CLC process. It was, more precisely, predicted that $KTi_8O_{16}$ would form under reducing conditions if the concentration of K is low and Ti high, which is expected on the inside of the ilmenite OC particles, in agreement with experimental observations.

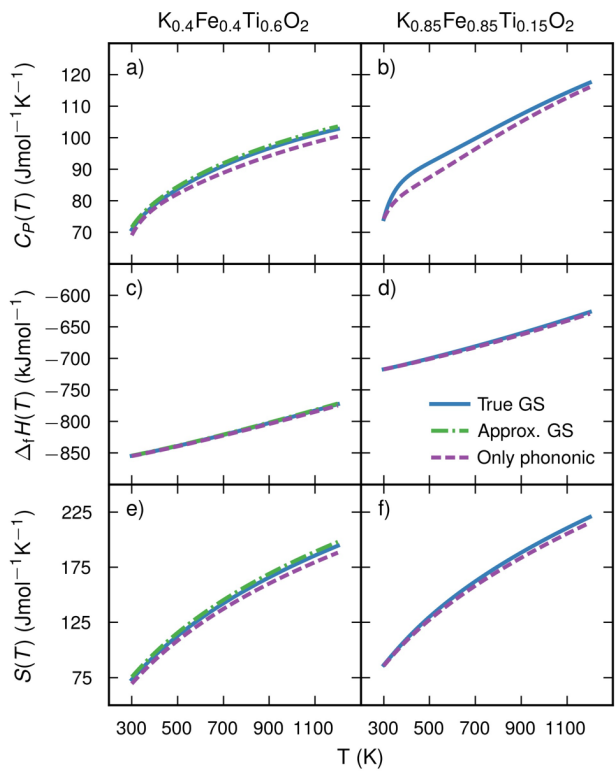

FIG. 4. Comparison between the heat capacity (a-b), enthalpy of formation (c-d), and entropy (e-f) obtained with (solid blue line) and without (dashed purple line) the contribution from the chemical disorder for $K_{0.4}Fe_{0.4}Ti_{0.6}O_2$ (a, c, e) and $K_{0.85}Fe_{0.85}Ti_{0.15}O_2$ (b, d, e). For the former compound, data for the approximate ground state (dash-dotted green line) is also shown.

When considering the entire ternary phase space of $Fe_2O_3$ –$K_2O$–$TiO_2$, the changes induced by the addition of the first-principles data are minor (see Figure 7). At a temperature of 950 °C and relative oxygen partial pressure of $10^{-14}$, these are limited to the portion closest to the $TiO_2$ corner ($TiO_2 \gtrsim 0.85$ and $Fe_3O_2 \lesssim 0.3$) To be precise, it forms together with ilmenite and first $K_2Ti_6O_{13}$, thereby replacing the titania spinel, followed by rutile as the proportion of Ti increases, and Fe decreases.

The other phases, more specifically $CaMn_{14}SiO_{24}$ and $Ca_{19}Mn_2(PO_4)_{14}$ as well as $KTi_8O_{16.5}$, $K_{0.4}Fe_{0.4}Ti_{0.6}O_2$, and $K_{0.85}Fe_{0.85}Ti_{0.15}O_2$, are not found to be stable under reasonable conditions. While it is true that these have been experimentally observed [8], specifically when analyzing the interaction between oxygen carriers and ash components in CLC of biomass, one should keep in mind that our calculations are based on several assumptions. Kinetics are, for instance, not taken into account, and the resulting predictions will, hence, not include any metastable materials.

## IV. CONCLUSIONS

This study clearly demonstrates that the thermodynamic properties of metal oxides can be efficiently estimated based on a combination of DFT and harmonic phonon calculations with a level of accuracy comparable to experimental measurements. The methodology is used for a range of different compounds with varying complexity. In the case of ilmenite ($FeTiO_3$), for which experimentally determined thermodynamic data is available in the FactSage databases, the estimated error is shown to be surprisingly small. When the first-principles data is included, the phase behavior changes for both Mn and Fe-based systems; the effect is the most prominent for K–Mn–O and K–Ti–O. As has been previously reported, this directly affects the simulation of processes involving CEs, such as CLC [8], which testifies to the potential of the method at hand.

The approach applied in this study is expected to not only be readily applicable to almost any metal oxide but also convenient for obtaining approximate values on the thermal properties of compounds that are not found in commercial or public databases. In addition, there might exist phases for which measurements are unreliable, for instance, because these are hard to synthesize in pure form. While such issues can be resolved via, e.g., deconvolution, the accuracy will most likely suffer. Significant discrepancies can also arise because of human errors, limitations of experimental techniques, as well as variations in the quality, structure and composition of the sample. Thus, it is not far-fetched to assume that there exist materials for which our methodology is more cost-effective and yields more reliable estimates than experiments.

Given the relative success of this study, we envision that thermodynamic properties could potentially be added to first-principles databases such as the OQMD [41, 42], the Materials Project [17], the aflowlib.org library [43, 44], and the Novel Materials Discovery (NOMAD) Laboratory [45]. While such an endeavor requires significant effort, the ever-increasing focus on computer-aided materials design and the growing number of high-throughput studies will, in our opinion, make it worthwhile.

## ACKNOWLEDGMENTS

This work was funded by the Swedish Research Council (2020-03487) and the computations were enabled by resources provided by the Swedish National Infrastructure for Computing (SNIC) at NSC and C3SE partially funded by the Swedish Research Council through grant agreement no. SNIC 2021/3-41, SNIC 2021/5-623, SNIC 2021/5-561, and SNIC 2022/5-156.

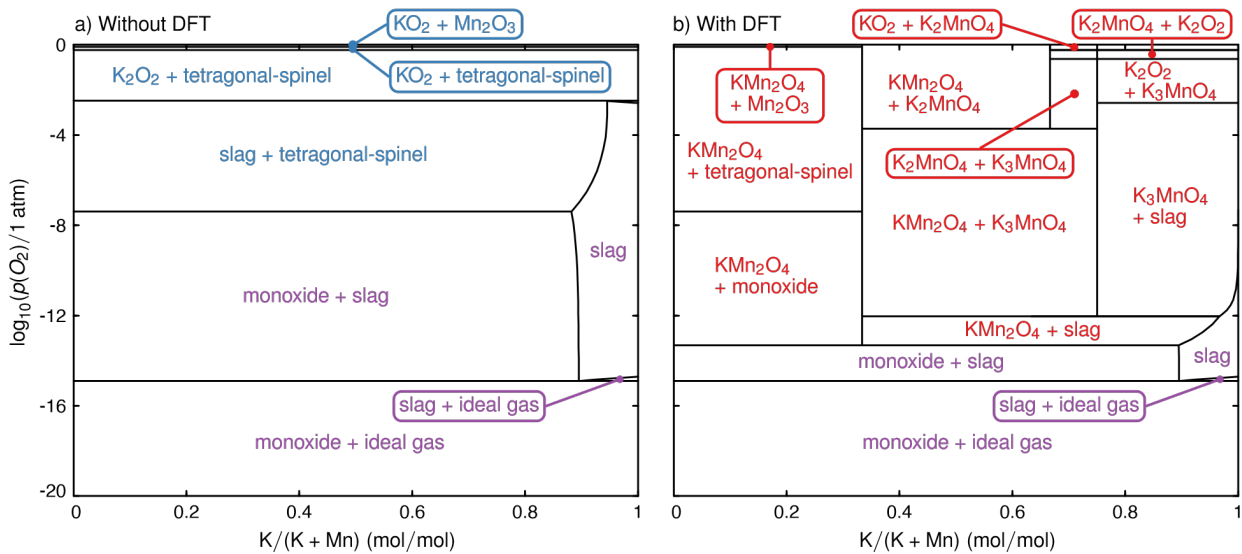

FIG. 5. Phase diagram for K–Mn–O showing $\log_{10}(p_{O2})$ versus K/(K+Mn) at 950°C and 1atm, generated using data from the FactPS and Toxid databases (a) as well as first-principles calculations (b). Common phases have purple labels, while those colored blue and red are only resent in panels a and b, respectively.

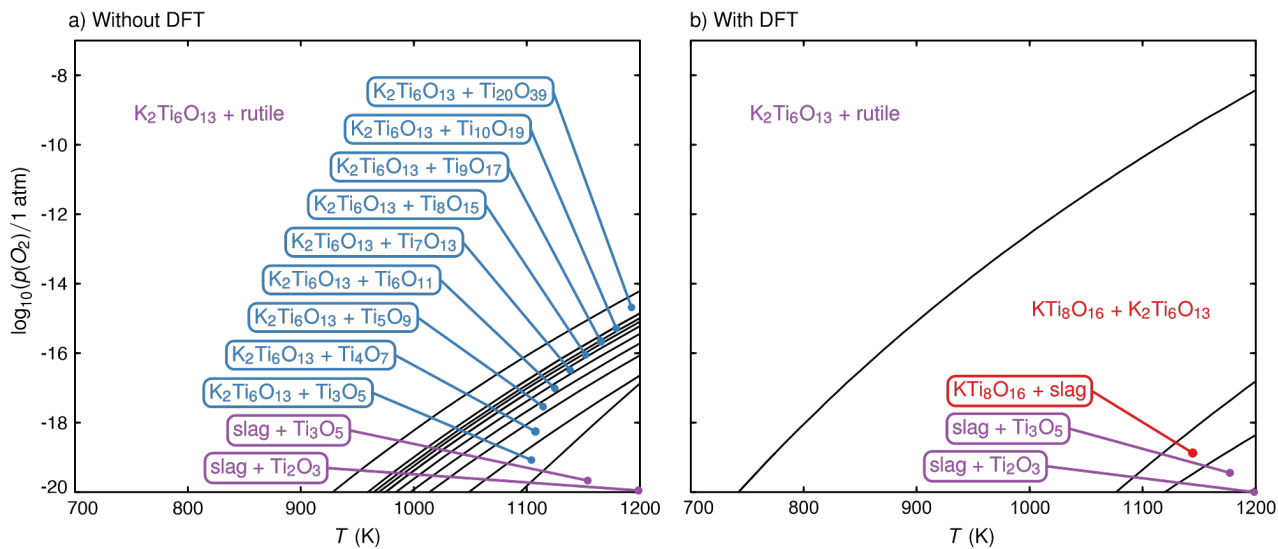

FIG. 6. Phase diagram for $K_2O$–$TiO_2$ showing $\log_{10}(p_{O2})$ versus $T$ at 1atm for a ratio of $K_2O/TiO_2 = 0.1$, generated using data from the FactPS nd FToxid databases (a) as well as first-principles calculations (b). Common phases have purple labels, while those colored blue and red are nly present in panels a and b, respectively.

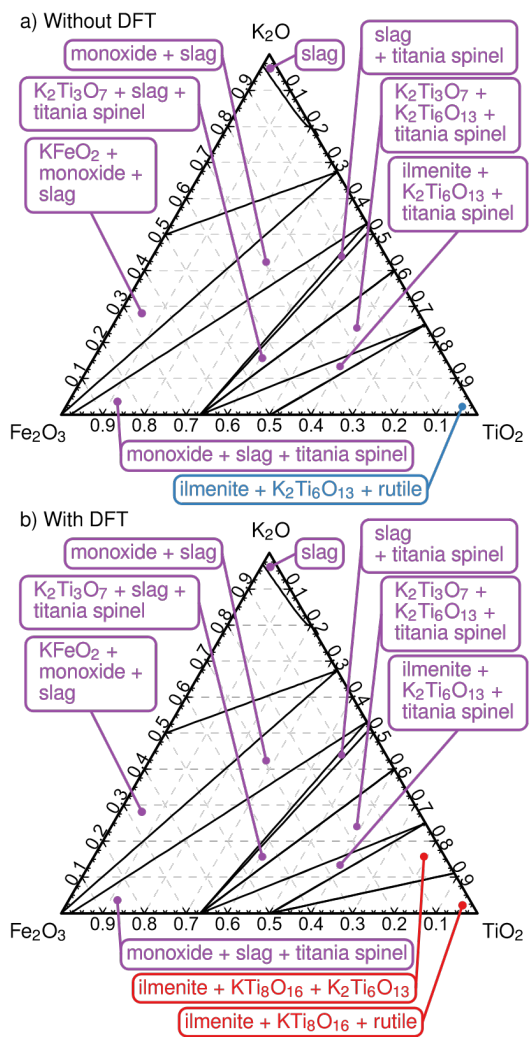

FIG. 7. Ternary phase diagram for $Fe_2O_3 - K_2O - TiO_2$ at 950℃, 1atm, and $p_{O_2} = 10^{-14}$ atm, generated using data from the FactPS and FToxid databases (a) as well as firstprinciples calculations (b). Common phases have purple labels, while those colored blue and red are only present in panels a and b, respectively.